\documentclass[aps,prd,twocolumn,amssymb,superscriptaddress,nofootinbib]{revtex4-2}
\usepackage{graphicx}
\usepackage{amsmath}
\usepackage{amsfonts}
\usepackage{mathrsfs}
\usepackage{orcidlink}
\usepackage[utf8]{inputenc}

\usepackage{hyperref}
\hypersetup{
  colorlinks=true,
  linkcolor=blue,
  citecolor=magenta,
  filecolor=magenta,
  urlcolor=blue
}

\def\beq{\begin{equation}}
\def\eeq{\end{equation}}
\def\L{\mathscr{L}}
\def\H{\mathcal{H}}
\def\MA{M_{\rm ADM}}
\def\QA{Q_{\rm ADM}}
\def\Qe{Q_{\rm e}}
\def\rh{r_{\rm h}}
\def\mh{m_{\rm h}}
\def\Th{T_{\rm h}}
\def\Ph{\Phi_{\rm h}}

\begin{document}

\title{Regular black holes do not violate the first law of thermodynamics}

\author{Bobir Toshmatov\,\orcidlink{0000-0002-7853-186X}}
\email{toshmatov@astrin.uz}
\affiliation{New Uzbekistan University, Movarounnahr street 1, Tashkent 100000, Uzbekistan}
\affiliation{Ulugh Beg Astronomical Institute, Astronomy street 33, Tashkent 100052, Uzbekistan}

\author{Bobomurat Ahmedov\,\orcidlink{0000-0002-1232-610X}}
\email{ahmedov@astrin.uz}
\affiliation{Institute of Theoretical Physics, National University of Uzbekistan, Tashkent 100174, Uzbekistan}
\affiliation{School of Physics, Harbin Institute of Technology, Harbin 150001, People's Republic of China}

\author{Nozima Isamadinova}
\email{n.isamadinova@newuu.uz}
\affiliation{New Uzbekistan University, Movarounnahr street 1, Tashkent 100000, Uzbekistan}

\author{Chengxun Yuan\,\orcidlink{0000-0002-2308-6703}}
\email{yuancx@hit.edu.cn}
\affiliation{School of Physics, Harbin Institute of Technology, Harbin 150001, China}

\begin{abstract}
Regular black holes in general relativity coupled to nonlinear electrodynamics have been reported to exhibit an apparent mismatch in their thermodynamics: the temperature obtained by differentiating the mass along the family of regular metrics differs from the Hawking temperature by a factor $\Xi$, and the first law seems to need a modified entropy or a corrected internal energy. We examine this question for an electrically charged family of regular black holes in $D\geq4$ dimensions whose metrics reduce to the Bardeen and Hayward geometries in four dimensions. The source is written as a Hamiltonian density $\H(P)$ with fixed couplings and no solution parameters. This theory has a two-parameter family of static solutions, of which the regular metrics form a one-parameter subfamily. We prove analytically that on the whole solution space the standard first law of thermodynamics holds with the Hawking temperature, the area entropy and the horizon potential, in agreement with the general first law for nonlinear electrodynamics. The factor $\Xi$ appears when the mass is differentiated along the regular metric family at fixed length parameter, a variation that moves through the space of theories and also changes the conserved charge. We show that $\Xi^{-1}$ is the Jacobian of this projection, equal to the fraction of the mass inside the horizon, give its general form for an arbitrary regular mass function, and show that $\Xi^{-1}d\MA$ is not a closed one-form on the two-parameter metric family, so this correction cannot be the differential of a local state function there. We also point out that the Legendre map between the $F$ and $P$ frames degenerates at a radius that lies outside the horizon of near-extremal solutions in $D\geq5$.
\end{abstract}

\maketitle

\section{Introduction}

It is well known that the simplest way to obtain a black hole solution without the central singularity is to couple general relativity (GR) to nonlinear electrodynamics (NED) \cite{Ayon-Beato98,Ayon-Beato00,Bronnikov:PRL:2000,Bronnikov:PRD:2001,Dymnikova:2004zc,Hayward:PhysRevLet:,Fan:PRD:2016}. The first law of black hole mechanics for GR coupled to NED is also well known. It was derived by Rasheed \cite{rasheed2003}, it follows from the covariant phase space method \cite{Zhang:2016ilt,Bokuli2021}, and if the Lagrangian of NED contains dimensionful parameters, then their variations enter the first law with their own conjugate potentials \cite{Zhang:2016ilt,Fan:PRD:2016,Bokuli2021}. In this law the temperature is the Hawking temperature $\Th$ and the entropy is given by the area of the horizon.

Nevertheless, in a large number of papers on the regular black holes in NED it is claimed that the first law does not hold in its standard form. The reason is the following. If the ADM mass of the regular black hole is written as a function of the horizon radius and of the parameter $Q$ which appears in the metric, and then differentiated at fixed $Q$, one obtains the temperature $T_{\rm thermo}$ which differs from $\Th$ by the factor $\Xi$. In order to restore the first law, some authors have modified the entropy, and some others have multiplied the variation of the mass by the correction factor and called the result the variation of a new internal energy \cite{Ma:2014qma,Breton:2004qa}. However, the reason for the mismatch is already contained in the papers cited above. When the regular metric is prescribed and the Lagrangian is derived from the field equations, the Lagrangian contains the mass and the charge of this particular solution. Therefore, the differentiation of the mass changes the theory itself, and the terms which are missing in the first law are exactly the variations of the parameters of the Lagrangian \cite{Zhang:2016ilt,Fan:PRD:2016,Bokuli2021}. Fan and Wang \cite{Fan:PRD:2016} have shown this explicitly for the four-dimensional case: they added the Schwarzschild mass to the solution, found the regular subset of the solutions, noted that for the electric solutions the Lagrangian $\L(F)$ is double-valued and one has to use the dual $P$ frame, and wrote the first law with the Hawking temperature and with the term corresponding to the variation of the parameter of the Lagrangian. Recently, Bokuli\'c, Juri\'c and Smoli\'c \cite{Bokulic:2025brf} have proved that the mass and the charge of the regular black hole in NED cannot be independent, and have shown how the parameters of the solution can be removed from the Bardeen Lagrangian by redefinitions.

Thus, we do not claim that the restoration of the first law at fixed couplings is a new principle. The aim of the present paper is to show the whole mechanism explicitly for a concrete family of black holes in arbitrary dimension $D\geq4$, and to add the following results which, to our knowledge, have not been given before: (i) the analytic proof of the full first law with the charge term on the whole two-parameter family of solutions of the fixed theory; (ii) the clear separation of the three parameter spaces which are usually mixed in the literature, and the restriction of the first law to the regular subfamily; (iii) the interpretation of $1/\Xi$ as the Jacobian of the projection of the extended first law onto the family of regular metrics, which is equal to the fraction of the ADM mass inside the horizon, and its general form $\Xi=(\partial\hat m(\rh;M,Q)/\partial M)^{-1}$ for an arbitrary regular mass function; (iv) the proof that $\Xi^{-1}d\MA$ is not a closed one-form on the family of regular metrics, i.e., the correction factor of Ref.~\cite{Ma:2014qma} cannot be the differential of a local internal energy there; and (v) the observation that the Legendre transformation between the $F$ and $P$ frames is degenerate outside the horizon of the near-extremal black holes in $D\geq5$. The family of black holes itself, its horizons, energy conditions and Smarr formula are presented in the companion paper \cite{companion}. In order to keep the present paper short, we do not reproduce the expressions derived there as separate equations. Where they are needed, we quote them in the text with reference to the companion paper, and only the quantities on which the present argument is based are given in full.

\section{The theory and its solutions}

Let us consider the static, spherically symmetric, electrically charged solutions of GR coupled to NED in $D$ dimensions with the line element $ds^2=-f\,dt^2+dr^2/f+r^2d\Omega_{D-2}^2$ and the electromagnetic potential $A=\varphi(r)dt$. For the electric field the usual Lagrangian $\L(F)$, where $F=F_{\mu\nu}F^{\mu\nu}$, is double-valued on the regular solution, since $F$ vanishes both at the center and at infinity \cite{Bronnikov:PRL:2000,Bronnikov17}. Therefore, we define the theory by the first-order action
\beq\label{action}
I=\frac{1}{16\pi}\int d^Dx\sqrt{-g}\left[R-2P^{\mu\nu}F_{\mu\nu}+\H(P)\right],
\eeq
where $F_{\mu\nu}=2\partial_{[\mu}A_{\nu]}$ and the antisymmetric tensor $P^{\mu\nu}$ are varied independently, and $P=P_{\mu\nu}P^{\mu\nu}$ \cite{Salazar:1987ap,Ayon-Beato98}. The variation with respect to $P^{\mu\nu}$ gives the constitutive relation $F_{\mu\nu}=\H_PP_{\mu\nu}$, the variation with respect to $A_\mu$ gives $\nabla_\mu P^{\mu\nu}=0$, and in the region where $\H_P\neq0$ and the map $P\mapsto F=P\H_P^2$ is locally invertible, the theory (\ref{action}) is equivalent to the usual one with $\L=2P\H_P-\H$, $P_{\mu\nu}=\L_FF_{\mu\nu}$ and $\H_P=1/\L_F$. For the static electric field one has $P<0$, and we use the action (\ref{action}) only in this domain. The energy density and the tangential pressure are given by $\rho=-\H/16\pi$ and $p_t=-\L/16\pi$, respectively.

In the present paper, we study the Hamiltonian density in the form
\beq\label{HP}
\H(P)=-\,\sigma\left[1+\left(\frac{\beta}{|P|}\right)^{\frac{\gamma}{2(D-2)}}\right]^{-\frac{\gamma+D-1}{\gamma}},\qquad P<0,
\eeq
where $\gamma>0$ is a dimensionless constant, and $\sigma$ and $\beta$ are two coupling constants with the dimension of length$^{-2}$. Note that (\ref{HP}) does not contain any parameter of the solution, in the same way as the Born-Infeld electrodynamics contains only its critical field. In the strong field limit, $|P|\to\infty$, the Hamiltonian density tends to the constant $-\sigma$, i.e., $\sigma$ plays the role of the vacuum energy of the core. In the weak field limit, $P\to0^-$, one has $\H\simeq-\sigma(|P|/\beta)^a$ with $a=(\gamma+D-1)/2(D-2)$. Therefore, the theory has the Maxwell limit, $\H\to P$, only if $a=1$ and $\sigma=\beta$, i.e., for $\gamma=D-3$. For $\gamma<D-3$ the derivative $\H_P$ diverges at $P=0$, so that $\H$ is smooth in the domain $P<0$, but not at the point $P=0$. We do not use (\ref{HP}) for $P>0$, where its formal extension would give the wrong sign of the Maxwell term. Since the normalization of the electromagnetic field is a matter of convention, the rescaling $P_{\mu\nu}\to cP_{\mu\nu}$ changes $\beta\to c^2\beta$ and does not change $\sigma$. Thus, $\beta$ fixes the unit of the charge and $\sigma$ is the only physical coupling constant of the theory. Nevertheless, below we keep $\sigma$ and $\beta$ as independent constants, because this two-parameter space of theories is the natural one for the extended first law.

The static electric solutions of the theory (\ref{HP}) are obtained by direct integration. The Gauss law gives $P=-2\QA^2/r^{2(D-2)}$, where $\QA$ is an integration constant, and the conserved electric charge is $\Qe=\Omega_{D-2}\QA/4\pi$, where $\Omega_{D-2}$ is the area of the unit $(D-2)$-dimensional sphere. It is convenient to replace $\QA$ by the length $\lambda=(2\QA^2/\beta)^{1/2(D-2)}$. Then $\beta/|P|=(r/\lambda)^{2(D-2)}$ and $\rho=\sigma/16\pi[1+(r/\lambda)^\gamma]^{(\gamma+D-1)/\gamma}$. Writing the metric function in the form $f=1-\hat m(r)/r^{D-3}$, the $tt$ component of the Einstein equations reads $\hat m'=16\pi\rho\,r^{D-2}/(D-2)$ and gives
\beq\label{mhat}
\hat m(r)=m_0+\frac{\sigma\,r^{D-1}}{(D-2)(D-1)\,[1+(r/\lambda)^\gamma]^{(D-1)/\gamma}}\ ,
\eeq
where $m_0$ is the second integration constant. The ADM mass of the solution is $\MA=\alpha\hat m(\infty)$, where $\alpha=(D-2)\Omega_{D-2}/16\pi$. Near the center the metric function behaves as $f\simeq1-m_0/r^{D-3}-\sigma r^2/(D-1)(D-2)$. Thus, the center is regular only for $m_0=0$, i.e., the regularity is the property of the solution, not of the theory. Here and below by the regular solution we mean the one with bounded curvature invariants at $r=0$ and with the de Sitter core. Since the first correction to $f$ near the center is of the order of $r^{2+\gamma}$, the metric is infinitely differentiable at the center only for even integer $\gamma$; the geodesic completeness is not discussed here.

It is important to distinguish the following three parameter spaces. (a) The solutions of the fixed theory $(\sigma,\beta,\gamma)$ form the two-parameter family with the parameters $(m_0,\QA)$; most of them have the singular center. (b) The regular solutions of the same theory form the one-parameter subfamily $m_0=0$, for which
\beq\label{regline}
\MA=\frac{\alpha\sigma}{(D-2)(D-1)}\left(\frac{2\QA^2}{\beta}\right)^{\frac{D-1}{2(D-2)}}\propto \Qe^{\frac{D-1}{D-2}}\ ,
\eeq
i.e., the mass is determined by the charge. In four dimensions this relation reads $\MA\propto\Qe^{3/2}$, and it has been proved in Ref.~\cite{Bokulic:2025brf} that it is unavoidable for the regular black holes in NED. (c) The two-parameter family of regular metrics which is obtained by varying $(M,Q)$ in Eq.~(\ref{metric}) given below; two members of this family belong, in general, to two different theories. Any statement about ``the first law of regular black holes'' depends on which of these three spaces is meant.

From (a) and (b) one can make the following remark on the dynamics, but only in the conditional form. Let the regular black hole of this theory absorb a charged particle and let the final state again belong to the static family (\ref{mhat}). Then, unless the changes of $\MA$ and $\Qe$ satisfy the tangent relation $\delta\MA=\frac{D-1}{D-2}(\MA/\Qe)\,\delta\Qe$ following from (\ref{regline}), the final state has $m_0\neq0$ and the singular center. Whether the real process ends in this family and what is radiated during the process are the questions of dynamics which we do not consider here.

\section{The family of regular metrics}

For $m_0=0$ Eq.~(\ref{mhat}) gives the metric function
\beq\label{metric}
f(r)=1-\frac{Mr^2}{(r^\gamma+Q^\gamma)^{(D-1)/\gamma}}\ ,
\eeq
with $Q=\lambda$ and $M=\sigma\lambda^{D-1}/(D-2)(D-1)$. Here $Q>0$ is the length parameter of the metric and $M$ has the dimension of length$^{D-3}$. The parameters $M$ and $Q$ are related to the coupling constants as
\beq\label{couplings}
\sigma=\frac{(D-2)(D-1)M}{Q^{D-1}}\ ,\qquad \beta=\frac{2\QA^2}{Q^{2(D-2)}}\ .
\eeq
We normalize the electromagnetic field in such a way that for $\gamma=D-3$ the weak field limit is exactly the Maxwell one, $\sigma=\beta$. Then $\beta=(\gamma+D-1)\sigma/2(D-2)$ for any $\gamma$, and the conserved charge of the solution (\ref{metric}) is $\QA^2=(D-1)(\gamma+D-1)MQ^{D-3}/4$; any other normalization multiplies $\QA$ by a constant. We emphasize that $Q$ is not the conserved charge, $\Qe^2\propto MQ^{D-3}$, and the variation of $M$ at fixed $Q$ changes $\Qe$. In four dimensions the cases $\gamma=2$ and $\gamma=3$ correspond to the Bardeen and Hayward metrics, respectively, and $\gamma=1$ corresponds to the solution of Ref.~\cite{Fan:PRD:2016}. Note that our electric solutions have the same metrics as these models, but not necessarily the same source: the Bardeen metric was originally interpreted as the field of the magnetic monopole \cite{Ayon-Beato00}, and for the Hayward metric the source was not specified \cite{Hayward:PhysRevLet:}. The Lagrangian density evaluated on the solution (\ref{metric}) and written as a function of $r$ has the form
\beq\label{L}
\L(r)=(D-1)MQ^\gamma\,\frac{(D-2)Q^\gamma-(\gamma+1)r^\gamma}{(r^\gamma+Q^\gamma)^{(2\gamma+D-1)/\gamma}}\ .
\eeq
At large $r$ one obtains $\L\propto|F|^{(\gamma+D-1)/2(\gamma+1)}$, i.e., the asymptotic electric branch is Maxwellian only for $\gamma=D-3$, while at the center, where $F\to0$ again, $\L\to\sigma$. This is the double-valuedness of $\L(F)$ mentioned above. Thus, the Bardeen and Hayward metrics have the asymptotically Maxwellian electric source in five and six dimensions, respectively, but not in four dimensions \cite{companion}.

From (\ref{metric}) the horizon radius is determined by $M=(\rh^\gamma+Q^\gamma)^{(D-1)/\gamma}/\rh^2$, the area entropy is $S=\Omega_{D-2}\rh^{D-2}/4$, and the Hawking temperature is
\beq\label{Th}
\Th=\frac{1}{4\pi\rh}\left(\frac{(D-1)\rh^\gamma}{\rh^\gamma+Q^\gamma}-2\right),
\eeq
which vanishes at $r_{\rm ext}=(2/(D-3))^{1/\gamma}Q$. Now let us consider $\MA(\rh,Q)$ as a function on the family of metrics (c) and differentiate it at fixed $Q$. One obtains $d\MA=T_{\rm thermo}dS+\mathcal{N}\,dQ$ with $\mathcal{N}=(D-1)Q^{\gamma-1}\MA/(\rh^\gamma+Q^\gamma)$ and
\beq\label{Xi}
\Xi\equiv\frac{(\partial\MA/\partial S)_Q}{\Th}=\frac{M}{\rh^{D-3}}=\frac{\MA}{\mh}\ ,\quad \mh=\alpha\rh^{D-3},
\eeq
where the ratio is understood as the algebraic identity between two functions of $(\rh,Q)$ which holds also in the extremal case, when both of them vanish, and $\mh$ is the generalized Misner-Sharp mass at the horizon. Since $\Xi>1$ for $Q>0$, this is not the first law with the Hawking temperature; the same happens in the models of Refs.~\cite{Ayon-Beato98,Bronnikov:PRL:2000,Dymnikova:2004zc}. Two remarks are in order. First, according to (\ref{couplings}) the above variation changes $\sigma$ and $\beta$, i.e., it moves through the space of theories. Second, it also changes the conserved charge. If one keeps $\Qe$ fixed instead of $Q$, one finds $(\partial\MA/\partial S)_{\Qe}=\Xi\Th\,(D-3)(\rh^\gamma+Q^\gamma)/[(D-3)\rh^\gamma+2(D-2)Q^\gamma]$, which can be both larger and smaller than $\Th$; for $D=4$, $\gamma=2$ and $\rh=2Q$ these two ratios are equal to $1.40$ and $0.87$. Thus, the value and even the sign of the mismatch depend on what is kept fixed, and neither of the two variations is the variation within one theory. For the general mass function $\hat m(r;M,Q)$ with $\hat m(\infty)=M$ the same calculation gives $\Xi=(\partial\hat m(\rh;M,Q)/\partial M)^{-1}$ at fixed $Q$; Eq.~(\ref{Xi}) is the particular case when $\hat m$ is proportional to $M$ with the profile independent of $M$.

\section{The first law at fixed couplings}

Now let us keep $(\sigma,\beta,\gamma)$ fixed and vary the integration constants $(m_0,\QA)$ of (\ref{mhat}), i.e., move in the space of solutions (a). We use the following assumptions: the static spherically symmetric metric with $g_{tt}g_{rr}=-1$, which follows from $T^t{}_t=T^r{}_r$ for NED; the Einstein gravity with the area entropy; and the energy density $\rho(r;\QA)$ which is fixed by the theory and does not depend on $m_0$. Integrating $\hat m'=16\pi\rho\,r^{D-2}/(D-2)$ from the horizon, where $\hat m(\rh)=\rh^{D-3}$, to infinity we get
\beq\label{Mint}
\MA=\mh+\Omega_{D-2}\int_{\rh}^\infty\rho(r;\QA)\,r^{D-2}\,dr\ ,
\eeq
i.e., $\MA$ is the function of $(\rh,\QA)$, or, equivalently, of $(S,\Qe)$. Let us differentiate it at fixed $\QA$. Using the expression $4\pi\Th=(D-3)/\rh-\hat m'(\rh)/\rh^{D-3}$ for the Hawking temperature of the metric $f=1-\hat m/r^{D-3}$ and $dS/d\rh=4\pi\alpha\rh^{D-3}$, one obtains
\beq\label{TdS}
\left(\frac{\partial\MA}{\partial\rh}\right)_{\QA}=\alpha\left[(D-3)\rh^{D-4}-\hat m'(\rh)\right]=\Th\frac{dS}{d\rh}\ .
\eeq
Now let us differentiate (\ref{Mint}) at fixed $\rh$. Since $\rho=-\H(P)/16\pi$ and $P=-2\QA^2/r^{2(D-2)}$, we have
\beq\label{drho}
\frac{\partial\rho}{\partial\QA}=\frac{\H_P\QA}{4\pi r^{2(D-2)}}=\frac{E(r)}{4\pi r^{D-2}}\ ,
\eeq
where $E=F_{tr}=\H_PP_{tr}=\H_P\QA/r^{D-2}$ is the electric field. Substituting (\ref{drho}) into (\ref{Mint}) and using $\Qe=\Omega_{D-2}\QA/4\pi$, we get
\beq\label{dQ}
\left(\frac{\partial\MA}{\partial\Qe}\right)_{S}=\int_{\rh}^\infty E\,dr=\Ph\ ,
\eeq
which is the electrostatic potential of the horizon, $\Ph=A_t(\rh)-A_t(\infty)$ in the gauge $E=-A_t'$, exactly as in the case of the Reissner-Nordstr\"om black hole. Combining (\ref{TdS}) and (\ref{dQ}) we obtain
\beq\label{firstlaw}
d\MA=\Th\,dS+\Ph\,d\Qe
\eeq
on the whole two-parameter family of solutions of the fixed theory, with the Hawking temperature and the standard area entropy. Note that no specific property of (\ref{HP}) has been used, i.e., Eqs.~(\ref{Mint})--(\ref{firstlaw}) hold for any $\H(P)$; they are the static electric case of the general first law of Refs.~\cite{rasheed2003,Zhang:2016ilt,Bokuli2021}. We have also checked (\ref{firstlaw}) numerically for $D=4$--$7$ and $\gamma=1$--$3$. On the regular subfamily the potential term can be written in the closed form $\Ph\Qe=\frac{D-1}{D-2}\big[\MA-\mh+\mh Q^\gamma/(\rh^\gamma+Q^\gamma)\big]$ \cite{companion}, which differs from $\mathcal{N}Q$, i.e., the coefficient $\mathcal{N}$ in Eq.~(\ref{Xi}) is not the potential of the horizon.

Two remarks should be made about the regular solutions. First, (\ref{firstlaw}) is the law on the space (a). At the regular black hole it is evaluated by comparison with the neighboring solutions which have $m_0\neq0$, i.e., the singular center, but the same structure of the exterior. This is the usual situation for the equilibrium first law, which never requires the neighboring solutions to have the same properties of the interior. Second, on the regular subfamily (b), where $S$ and $\Qe$ are not independent, Eq.~(\ref{firstlaw}) reduces to the single identity $\Th\,dS+\Ph\,d\Qe=\frac{D-1}{D-2}(\MA/\Qe)\,d\Qe$ along the curve (\ref{regline}), and there is no independent derivative $(\partial\MA/\partial S)_{\Qe}$ within (b) at all. The derivative in (\ref{TdS}) is the derivative in the larger space (a), and the same is true for the thermodynamic temperature of the regular black hole.

\section{The meaning of the factor $\Xi$}

Since on the family of metrics (c) the coupling constants are the functions of $M$ and $Q$, the natural extension of (\ref{firstlaw}) to this family is to consider the couplings as the thermodynamic variables, as it is done for the cosmological constant \cite{Kastor:2009wy} and for the parameters of NED \cite{Zhang:2016ilt,Fan:PRD:2016,Gulin:2017ycu,Bokuli2021}:
\beq\label{extended}
d\MA=\Th\,dS+\Ph\,d\Qe+\Psi_\sigma d\sigma+\Psi_\beta d\beta\ ,
\eeq
where $\Psi_\sigma$ and $\Psi_\beta$ are the derivatives of $\MA$ at fixed $S$ and $\Qe$ in the two-parameter space of theories (\ref{HP}). Both of them follow from (\ref{Mint}). Since $\rho$ is proportional to $\sigma$, one has $\sigma\Psi_\sigma=\MA-\mh$, which is the energy of the field outside the horizon. Since $\rho$ depends on $\beta$ and $\QA$ only through $\lambda\propto(\QA^2/\beta)^{1/2(D-2)}$, one has $\beta\partial_\beta=-\frac12\QA\partial_{\QA}$ and, therefore, $\beta\Psi_\beta=-\Ph\Qe/2$. Now let us restrict (\ref{extended}) to the family (c) with our normalization. From (\ref{couplings}) we have
\begin{align}
\frac{d\sigma}{\sigma}&=\frac{d\beta}{\beta}=\frac{dM}{M}-(D-1)\frac{dQ}{Q}\ ,\nonumber\\
2\frac{d\Qe}{\Qe}&=\frac{dM}{M}+(D-3)\frac{dQ}{Q}\ .\label{pullback}
\end{align}
Substituting these relations and the closed form of $\Ph\Qe$ into (\ref{extended}), after a short calculation we obtain
\begin{align}
\frac{\mh}{\MA}\,d\MA&=\Th\,dS+\Phi\,dQ\ ,\nonumber\\
\Phi&=\frac{(D-1)\alpha\,\rh^{D-3}Q^{\gamma-1}}{\rh^\gamma+Q^\gamma}=\frac{\mathcal{N}}{\Xi}\ .\label{projection}
\end{align}
Equation (\ref{projection}) is nothing but Eq.~(\ref{Xi}) divided by $\Xi$, i.e., these are the same statements. The prefactor $\mh/\MA=1/\Xi$ is exactly the correction factor which Ma and Zhao \cite{Ma:2014qma} put in front of $d\MA$ and interpret as the variation of the corrected internal energy. Now we see its meaning: it is the Jacobian which remains when the extended first law in the space of theories is projected onto the family of regular metrics, and it is equal to the fraction of the ADM mass inside the horizon. If $\Xi^{-1}d\MA$ were the differential of some function $U(\rh,Q)$, the one-form on the right-hand side of (\ref{projection}) would be closed. However, the direct calculation gives
\beq\label{nonexact}
\frac{\partial\Phi}{\partial\rh}-\frac{\partial}{\partial Q}\!\left(\Th\frac{dS}{d\rh}\right)=\frac{(D-1)(D-3)\alpha Q^{\gamma-1}\rh^{D-4}}{\rh^\gamma+Q^\gamma}\ ,
\eeq
which is strictly positive for $Q>0$ and $D\geq4$. Thus, the precise statement is the following: on the two-dimensional family of regular metrics (c) there is no local state function $U$ with $dU=\Xi^{-1}d\MA$ in any open region. Of course, along a one-dimensional curve this one-form can be integrated; the horizon mass $\mh$ is a state function, but $d\mh\neq\Xi^{-1}d\MA$; and the energy function defined together with other work terms is not excluded by (\ref{nonexact}). What is excluded is the particular proposal of Ref.~\cite{Ma:2014qma}, in which $\Xi^{-1}d\MA$ itself is the differential of the energy. At the same time, $d\MA$ is exact and its first law (\ref{firstlaw}) does not need any correction.

\section{Remark on the matter theory}

The $P$ frame makes the thermodynamic calculation simple, but the single-valuedness of $\H(P)$ does not guarantee that the matter theory is well defined. The map $P\mapsto F=P\H_P^2$ has the derivative $dF/dP=\H_P(\H_P+2P\H_{PP})$, and on the solutions of the theory (\ref{HP}), with $x=(r/\lambda)^\gamma$, one has
\beq\label{degen}
1+\frac{2P\H_{PP}}{\H_P}=\frac{(\gamma+1)x-(\gamma+D-2)}{(D-2)(1+x)}\ ,
\eeq
which vanishes at $r_{\rm b}=\lambda\,[(\gamma+D-2)/(\gamma+1)]^{1/\gamma}$. This is the branch point of $\L(F)$, where $\L_{FF}$ diverges and the effective metric which governs the propagation of the electromagnetic perturbations is singular \cite{Bronnikov:PRD:2001,Novello00}. Comparing with $r_{\rm ext}$ given after (\ref{Th}), we find that $r_{\rm b}<r_{\rm ext}\leq\rh$ for $D=4$, i.e., in four dimensions the degeneracy is always hidden behind the horizon. However, $r_{\rm b}>r_{\rm ext}$ for any $\gamma>0$ if $D\geq5$, i.e., in five and more dimensions there is a range of near-extremal black holes for which the degeneracy is located outside the event horizon \cite{companion}. The identities (\ref{firstlaw})--(\ref{nonexact}) are the statements about the static backgrounds and they are not affected by this. The question whether these backgrounds are acceptable solutions of the field theory with the well-posed dynamics requires the analysis of the characteristics of (\ref{action}) across $r_{\rm b}$, which is beyond the scope of the present paper. Therefore, the results of this paper should be understood as the identities for the backgrounds, not as the statement about the dynamical viability.

\section{Conclusion}

In the present paper we have shown the following. For the family of theories (\ref{HP}) in any dimension $D\geq4$ the standard first law (\ref{firstlaw}) holds on the whole two-parameter family of solutions, with the Hawking temperature, the area entropy and the electrostatic potential of the horizon. This follows from the simple analytic calculation (\ref{Mint})--(\ref{dQ}), which uses only the assumptions listed there and is the static electric case of the general first law \cite{rasheed2003,Zhang:2016ilt,Bokuli2021}. The regular black holes form the one-parameter subfamily of this space, in which the mass is determined by the charge, and the first law at the regular black hole is evaluated by comparison with its singular neighbors. The temperature mismatch reported in the literature is the derivative of the mass along the two-parameter family of regular metrics at fixed length parameter of the metric. This family is not the space of solutions of one theory, and this variation changes also the conserved charge. The projection of the extended first law with the coupling terms onto this family reproduces the mismatch exactly, with $\Xi^{-1}=\mh/\MA$ as the Jacobian, and shows that $\Xi^{-1}d\MA$ is not a closed one-form there. The same construction can be repeated for any regular metric which is obtained by prescribing $f$ and deriving the source, if the source can be rewritten with fixed couplings and the above assumptions are satisfied; in this case the general form of the factor is $\Xi=(\partial\hat m(\rh;M,Q)/\partial M)^{-1}$, and we do not claim that it is universal beyond this. Finally, the Hamiltonian formulation removes the double-valuedness of $\L(F)$ from the thermodynamic calculation, but not the Legendre degeneracy at $r_{\rm b}$, which in $D\geq5$ can be located outside the horizon. This problem is common for all electric regular solutions and deserves a separate study.

\begin{acknowledgments}
This research was partially supported by the Ministry of Higher Education, Science, and Innovation of the Republic of Uzbekistan under Grants No. FL-9524114991 and No. FL-10425067111. B.A. was supported by the National Natural Science Foundation of China (NSFC) under Grant No. U2541210.
\end{acknowledgments}

\bibliography{references}

\end{document}